\documentclass{llncs}

\usepackage{mathtools}

\usepackage{lscape}
\usepackage{rotating}
\usepackage{adjustbox}
\usepackage{caption}
\usepackage{listings}

\usepackage{wrapfig}
\usepackage{array}
\usepackage{graphicx}

\usepackage{makebox}

\usepackage{algorithm}
\usepackage{algorithmicx}
\usepackage{algpseudocode}

\usepackage{fancyhdr} 
\usepackage{booktabs}
\usepackage{subfigure}
\usepackage{multirow}

\usepackage{tikz}
\usetikzlibrary{arrows.meta}
\usepackage{framed} 

\usepackage{amsmath}
\usepackage{amssymb}

\usepackage{lscape}
\usepackage{appendix}

\usepackage{comment}

\usepackage{setspace}
\usepackage{ntheorem}

\usepackage[
	n,
	operators,
	advantage,
	sets,
	adversary,
	landau,
	probability,
	notions,	
	logic,
	ff,
	mm,
	primitives,
	events,
	complexity,
	asymptotics,
	keys]{cryptocode}
	
\usepackage{dashbox}
\usepackage{tikz}
\usepackage{enumerate}

\usepackage{enumitem}
\usepackage{marvosym}

\begin{document}

\title{BDTS: Blockchain-based Data Trading System}

\renewcommand\rightmark{BDTS: Blockchain-based Data Trading System}
\newcommand{\romannum}[1]{\romannumeral #1}






\author{Erya Jiang\inst{1}, Bo Qin\inst{1}, Qin Wang\inst{2}, Qianhong Wu\inst{3}, Sanxi Li\inst{1}, \\Wenchang Shi\inst{1}, Yingxin Bi\inst{1}, and Wenyi Tang\inst{4}}
\titlerunning{BDTS: Blockchain-based Data Trading System}

\institute{
\textit{Renmin University of China, China}\\
\and
\textit{CSIRO Data61, Australia} \\
\and
\textit{Beihang University, China} \\
\and
\textit{University of Notre Dame, America} \\
}


\maketitle           
\begin{abstract}
Trading data through blockchain platforms is hard to achieve \textit{fair exchange}. Reasons come from two folds: Firstly, guaranteeing fairness between sellers and consumers is a challenging task as the deception of any participating parties is risk-free. This leads to the second issue where judging the behavior of data executors (such as cloud service providers) among distrustful parties is impractical in the context of traditional trading protocols. To fill the gaps, in this paper, we present a \underline{b}lockchain-based \underline{d}ata 
\underline{t}rading \underline{s}ystem, named BDTS. BDTS implements a fair-exchange protocol in which benign behaviors can get rewarded while dishonest behaviors will be punished. Our scheme requires the seller to provide consumers with the correct encryption keys for proper execution and encourage a rational data executor to behave faithfully for maximum benefits from rewards. We analyze the strategies of consumers, sellers, and dealers in the trading game and point out that everyone should be honest about their interests so that the game will reach Nash equilibrium. Evaluations prove efficiency and practicability. 

\end{abstract}


\section{Introduction}

Data has risen to a new factor of production alongside traditional factors such as land, labor, capital and technology. 
Consumers, sellers, and data trading intermediaries together form a thriving data trading ecosystem, in which the consumer has to pay a fortune to the seller for acquisition, the seller could make some profits by providing the appropriate data, and data trading intermediaries earn agent fees between sellers and consumers.
However, such a high density of centralization is likely to be the weak spot to be attacked. On the one hand, any participating roles may act maliciously in the unsupervised system. The sellers may provide fake data for profits as they may not own data as they claimed. The consumer may refuse to pay after receiving the appropriate data. The intermediaries such as cloud service providers may manipulate the stored data without permission from users \cite{wang2011toward}\cite{zhu2011dynamic}.
On the other hand, relying on centralized servers confronts heavy communication and security pressure, greatly constraining the efficiency of the entire trading system. A single-point service provider may lead to unpleasant downloading experiences due to its limited resources, such as bandwidth, which may fail to cope with the environment of large-scale downloading. These challenges lead to the following questions,

\smallskip
\textit{Is it possible to propose a protocol in the data trading system with guaranteed fairness for all parties without significantly compromising efficiency?}
\smallskip


Traditional solutions using cryptography and relying on trusted third parties (TTP) \cite{kupccu2010usable}\cite{micali2003simple} lack practical significance because finding such a TTP
is reckon hard in practice.
Instead of using \textit{gradual release} method\footnote{The \textit{gradual release} method means each party in the game takes turns to release a small part of the secret. Once a party is detected doing evil, other parties can stop immediately and recover the desired output with similar computational resources.} \cite{blum1983exchange}\cite{pinkas2003fair}, many solutions \cite{dziembowski2018fairswap}\cite{he2021fair}\cite{shin2017t}\cite{choudhuri2017fairness}\cite{kiayias2016fair}\cite{eckey2020optiswap} have been proposed by leveraging blockchain technologies \cite{wood2014ethereum} for better decentralization. Blockchain provides a public bulletin board \cite{li2022smart} for every participating party with persistent evidence. A normal operating blockchain platform greatly reduces the risk of being attacked like a single-point failure or compromised by adversaries. Self-executing smart contracts always act benign and follow agreed principles, with transparency and accountability.

Based on such investigations, we adopt the blockchain technique as our baseline solution, with smart contracts acting as data executors. Specifically, we implement our scheme with strict logic of fair exchange on the Hyperledger blockchain.\cite{androulaki2018hyperledger} Overall, implementing a data trading system with both exchange fairness and efficiency for real usage is the key task in this paper. To fill the gap, our \textit{\textbf{contributions}} are as follows.

\begin{itemize}
\item We purpose BDTS, an innovative data trading system based on blockchain (\textcolor{magenta}{Sec.\ref{sec-archi}}\&\textcolor{magenta}{\ref{sec-system}}). The proposed scheme realizes the exchange fairness for \textit{all} participated parties, namely, \textit{consumer}, \textit{seller} and \textit{service provider}. Each party has to obey the rules, and benign actors can fairly obtain incentivized rewards. Every data can only be sold once since each transaction is unique in the blockchain systems. Notably, we use the uniqueness index mechanism\cite{chen2017bootstrapping} and compare Merkle roots of different data to prevent someone from reselling data purchased from others.

\item We prove the security of our scheme majorly from the economical side based on game theory  (\textcolor{magenta}{Sec.\ref{sec-security}}). Our proof simulates the behaviors of different parties, which is an effective hint to show actual reflection towards conflicts as well as real action affected by competitive participants. The proofs demonstrate that our game reaches \textit{the subgame perform equilibrium(SPE)} \cite{fang2021introduction}\cite{moore1988subgame}.

\item We implement our scheme on the Hyperledager Fabric blockchain platform with comprehensive evaluations (\textcolor{magenta}{Sec.\ref{sec-efficiency}}). Experimental results prove the efficiency and practicability. Compared to existing solutions with complex crypto-algorithms (e.g. zero-knowledge proof), our scheme is sufficiently fast for lightweight deceives.

\end{itemize}

\section{Related Work}
\label{sec-rw}

In this section, we provide related primitives surrounding \textit{fair exchange} protocols and \textit{blockchain}s. We compared the differences and main pros and cons between the references in Table \ref{tab:ref}.

\begin{wraptable}[5]{r}{0.65\textwidth}
\renewcommand\arraystretch{1.1}
\scriptsize
\vspace{-0.5in}
\setlength{\abovecaptionskip}{-0.3cm}
\caption{Reference Summary}\label{tab:ref}
\begin{center}
\begin{tabular}{r|cccccc}
  \toprule
  \rule{0pt}{3pt}\makebox[0.03\textwidth][r]{\textbf{Reference}}
  &\makebox[0.04\textwidth][c]{\cite{jung2017accounttrade}}
  &\makebox[0.04\textwidth][c]{\cite{chen2017bootstrapping}}
  &\makebox[0.11\textwidth][c]{\cite{dai2019sdte}\cite{li2020accountable}\cite{zhou2018distributed}}
  &\makebox[0.03\textwidth][c]{\cite{dziembowski2018fairswap}}
  &\makebox[0.06\textwidth][c]{\cite{he2021fair}\cite{eckey2020optiswap}}
  &\makebox[0.1\textwidth][c]{This work}\\
  
  \cmidrule(l){1-7}
  {\textbf{Decentralization}}
  &\checkmark
  &\checkmark
  &\checkmark
  &\checkmark
  &\checkmark
  &\checkmark\\

  {\textbf{Fairness}}
  &$\times$
  &$\times$
  &$\times$
  &\checkmark
  &\checkmark
  &\checkmark\\
  
  {\textbf{Prevent Resale}}
  &$\times$
  &\checkmark
  &$\times$
  &$\times$
  &$\times$
  &\checkmark\\
  
  {\textbf{Multimedia Data}}
  &\checkmark
  &$\times$
  &$\times$
  &$\times$
  &\checkmark
  &\checkmark\\
  
 \bottomrule
\end{tabular}
\end{center}
\vspace{0.2in}
\end{wraptable}
\noindent\textbf{Blockchain in Trading Systems.} Due to its non-repudiation, non-equivocation, and non-frameability, blockchain has been widely used in trading systems~\cite{li2022smart}. Jung et al. 
\cite{jung2017accounttrade} propose AccountTrade, an accountable trading system between customers who distrust each other. Any misbehaving consumer can be detected and punished by using book-keeping abilities. Chen et al.~\cite{chen2017bootstrapping} design an offline digital content trading system. If a dispute occurs, the arbitration institution will conduct it.  Dai et al.~\cite{dai2019sdte} propose SDTE, a trading system that protects data and prevents analysis code from leakage. They employ a trusted execution environment (TEE) to protect data in an isolated area at the hardware level. Similarly, Li et al.~\cite{li2020accountable} leverage the TEE-assisted smart contract to trace the evidence of investigators' actions. Automatic executions enable warrant execution accountability with the help of TEE. Zhou et al.~\cite{zhou2018distributed} introduce a data trading system that prevents index data leakage where participants exchange data via smart contracts. 
These solutions rely on blockchain to create persistent evidence and act as a transparent authority to solve disputes. 
However, they merely perform effectively in trading the \textit{text} data, rather than data cast in streaming channels such as TV shows and films, which are costly. The fairness issue has neither been seriously discussed. 

\smallskip
\noindent\textbf{Fair Exchanges using Blockchain.} Traditional ways of promoting fair exchange across distrustful parties rely on trusted third parties because they can monitor the activities of participants, judging whether they have faithfully behaved. However, centralization is the major hurdle. Blockchain can perfectly replace the role of TTP. The behaviors of involved parties are transparently recorded on-chain, avoiding any types of cheating and compromising. Meanwhile, a predefined incentive model can be automatically operated by smart contracts, guaranteeing that each participant can be rewarded according to their contributions.  Dziembowski et al. \cite{dziembowski2018fairswap} propose Fairswap, utilizing the smart contract to guarantee fairness. The contract plays the role of an external judge to resolve the disagreement. He et al. \cite{he2021fair} propose a fair content delivery scheme by using blockchain. They scrutinize the concepts of exchange fairness and delivery fairness during the trades. Eckey et al. \cite{eckey2020optiswap} propose a smart contract-based fair exchange protocol with an optimistic mode. They maximally decrease the interaction between different parties. Janin et al. \cite{janin2020filebounty} present FileBounty, a fair protocol using the smart contract. The scheme ensures a buyer purchases data at an agreed price without compromising content integrity. Besides, blockchains are further applied to 
multi-party computations of trading systems \cite{shin2017t}\cite{choudhuri2017fairness}\cite{kiayias2016blockchain}.

\section{Architecture and Security model}
\label{sec-archi}


\noindent\textbf{Entities.} First of all, we clarify the participating roles in our scheme. A total of three types of entities are involved: \textit{consumer (CM)}, \textit{seller (SL)}, and \textit{service provider (SP)}\footnote{Service providers generally act as the role of centralized authorities such as dealers and agencies in a traditional fair exchange protocol.}. Consumers pay for data and downloading service with cryptocurrencies such as Ether. Sellers provide encrypted data as well as expose the segment of divided data when necessary to guarantee correctness. Service providers take tasks of storage and download services, and any participant who stores encrypted data can be regarded as a service provider. Miners and other entities participating in systems are omitted as they are out of scope.

\smallskip
\noindent\textbf{Architecture.}  We design a novel data trading ecosystem that builds on the top of the blockchain platform. A typical workflow in BDTS is that: \textit{Sellers upload their encrypted data and description to service providers. Service providers store received data and establish download services. Consumers decide which pieces of data to purchase based on related descriptions. At last, the consumer downloads from service providers and pays the negotiated prices to sellers and service providers}. Our fair exchange scheme is used to ensure every participant can exchange data with payments without cheating and sudden denial. 

\noindent\hangindent 1em \textit{\textbf{Data upload.}} The seller first sends his description of data to the blockchain. Description and other information such as the Merkle root of data would be recorded by blockchain. Here, the seller must broadcast the Merkle root and they are demanded to expose a certain number of plaintext data pieces. Service providers can decide whether they are going to store it by the stated information. At the time, the seller waits for the decision from the service providers. If a service provider decides to store encrypted data to earn future downloading fees, he first sends his information to the blockchain. The seller will post encrypted data to the service provider and the service provider starts to store the data. Notably, the seller can also become a self-maintained service provider if he can build up similar basic services.

\noindent\hangindent 1em \textit{\textbf{Data download.}} The consumer decides to download or not according to the description and exposed parts provided by the seller. Before downloading, the consumer should first store enough tokens on the smart contract. Then, the consumer sends a request for data from service providers. Service providers will send it to the consumer after encrypting the data with the private key. For security and efficiency, these processes will be executed via smart contracts, except for data encryption and downloading. 

\noindent\hangindent 1em \textit{\textbf{Decryption and appealing.}} The consumer should pay for data and get the corresponding decryption key. The service provider and seller will provide their decryption key separately. The decryption key is broadcast through the blockchain so that it cannot be tampered with. The consumer can appeal whether it is due to the receipt of a false decryption key or the verification finds that the data has been falsified or fabricated. The smart contract will arbitrate based on evidence provided by the consumer.

\smallskip
\noindent\textbf{Security Assumption.} We have three basic security assumptions.
\textit{(i) The blockchain itself is assumed to be safe.} Our scheme operates on a safe blockchain model with well-promised \textit{liveness} and \textit{safety} \cite{garay2015bitcoin}. Meanwhile, miners are considered to be honest but \textit{curious}: they will execute smart contracts correctly but may be curious about the plaintext recorded on-chain. 
\textit{(ii) The basic crypto-related algorithm is safe.} This assumption indicates that the encryption and decryption algorithm will not suffer major attacks that may destroy the system. Specifically, AES and the elliptic curve, used for asymmetric encryption algorithms, are sufficiently supposed to be safe in cryptography. 
\textit{(iii) Participants in this scheme are rational.} As the assumption of game theory, all players (consumer, seller, and service provider) are assumed to be rational: these three types of players will act honestly but still pursue profits within the legal scope. 

\smallskip
\noindent\textbf{Security Model.}\label{subsec-secritymodel}  We dive into the strategies of each party.
\noindent\hangindent 1em \textit{\textbf{Seller}} intend to obtain more payment by selling their data. In our scheme, a seller needs to provide mainly three sectors: \textit{data}, \textit{description}, and \textit{decryption-key}. To earn profits, a seller would claim the data is popular and deserved to be downloaded, but he may provide fake data. 
The exchange is deemed as \textit{fair} if consumers obtain authentic data that is matched with claimed descriptions. Then, the seller can receive rewards. Encryption is another component provided by the seller. Only the correct decryption key can decrypt the encrypted data, whereas the false one cannot. In summary, there are four potential strategies for sellers: a) \textit{matched data (towards corresponding description) and matched key (towards data)}, b) \textit{matched data and non-matched key}, c) \textit{non-matched data and matched key}, and d) \textit{non-matched data and non-matched key}. 

\noindent\hangindent 1em \textit{\textbf{Consumer}} intend to exchange their money for data and downloading services. Downloading ciphertext and decrypting it to gain plaintext is in their favor. Consumers provide related fees in our scheme and then download encrypted data from service providers who store the uploaded data. To earn profits, they intend to obtain data without paying for it or paying less than expected. Paying the full price as expected for data is a sub-optimal choice. The payment of consumers can be divided into two parts: paying the seller for the decryption key and paying service providers for the downloading service. There are four strategies for consumers: a) \textit{pay enough for sellers}, b) \textit{pay less for sellers}, c) \textit{pay enough for service providers}, and d) \textit{pay less for service providers}.

\noindent\hangindent 1em \textit{\textbf{Service providers}} intend to provide the downloading service and earn profits. Service providers are like platforms, by storing as much data as possible and offering download service, they can ultimately attract clout and make a profit from the download fees. For uploading, service providers can choose whether to store data or not. Here, a seller can act as a service provider if he provides similar services of storage and download. For downloading, service providers will provide encrypted data and the corresponding decryption key. The strategies for service providers are listed as follows:  a) \textit{authentic correct data and matched key},  b) \textit{authentic data and non-matched key},  c) \textit{fake data and matched key},  and d) \textit{fake data and non-matched key}. The first two need the premise of storing the seller's data. 

\smallskip
\noindent\textbf{Strategy Assumption.}  For security, an ideal strategy for the system is to reach a Nash equilibrium for all participants: sellers adopt the \textit{correct data and matched key} strategy, consumers adopt the \textit{pay enough for sellers, paying enough for service providers} and service providers who provide storing services adopt the \textit{authentic correct data and matched key} strategy (discussed in \textcolor{magenta}{Sec.\ref{sec-security}}).

\section{The BDTS Scheme}
\label{sec-system}

In this section, we provide the concrete construction. To achieve security goals as discussed, we propose our blockchain-based trading system, called BDTS. It includes four stages: \textit{contract deployment}, \textit{encrypted data uploading}, \textit{data downloading}, and \textit{decryption and appealing}. Our scheme involves three types of contracts. Here, we omit procedures such as signature verification and block mining because they are known as common sense.

\smallskip
\noindent\textbf{Module Design.}
The system contains three types of smart contracts: seller-service provider matching contract (SSMC), service provider-consumer matching contract (SCMC), and consumer payment contract (CPC). 
Table \ref{tab:notation} outlines the notation used in the module description.

\begin{table}[!hbt]
\vspace{-0.4in}
\setlength{\abovecaptionskip}{-0.2cm}
\caption{Notation}\label{tab:notation}
\begin{center}
\resizebox{0.95\linewidth}{!}{
\begin{tabular}{cl} 
  \toprule
  \textbf{Notation} &   \textbf{Description}\\
  \midrule
  \textit{SL,SP,CM} &   seller/service provider/consumer\\
  $K_{role}$        &   the key of symmetric encryption algorithm of role\\
  $Data_i$          &   the \textit{i}-th unit of data plaintext in binary form\\
  $D_i$             &   $Enc^{AES}_{K_i}(Data_i)$,the \textit{i}-th unit of data encrypted by seller\\
  $DD_i$             &   $Enc^{AES}_{K_{sp}}(D_i)$,encrypted $D_i$ by service provider\\
  $Pub_{role}$,$Pri_{role}$
                    &   the public/private key of asymmetric encryption algorithm of role\\
  $A_{role}$        &   Ethereum address of role\\
  $IP_{role}$       &   IP address of role\\
  $ID_{data}$       &   the data ID, index of data in SSMC\\
  $Desc$            &   the seller’s description of data\\
  $\mathsf{Mtree}(·)$,$\mathsf{Mproof}(·)$,$\mathsf{Mvrfy}(·)$                            &   the Merkle tree algorithms\\
  $Tkn_{role}$        &   the token sent by role\\
  $Price$           &   the price of entire data\\
  $Unit\, Price$      &   downloading price for each unit\\
 \bottomrule
\end{tabular}
}
\end{center}
\vspace{-0.4in}
\end{table}
\noindent\hangindent 1em \textit{\textbf{Seller-service provider matching contract.}}  SSMC records the description and the Merkle root of data. The seller is required to broadcast certain parts of data and the index of these parts should be randomly generated by the blockchain. Notably, these indexes cannot be changed once they have been identified. Last, SSMC matches service providers for every seller.
   
\noindent\hangindent 1em \textit{\textbf{Service provider-consumer matching contract.}}  SCMC helps consumers and service providers reach an agreement. It receives and stores the consumers’ data, including required data and related information. The contract requires consumers to send payment. Then, the payment is sent to CPC.

\noindent\hangindent 1em \textit{\textbf{Consumer payment contract.}} CPC works to command consumers to pay for data and command sellers to provide the decryption key. It achieves a fair exchange between the decryption key (data downloading) and payment.

\smallskip
\noindent\textbf{Encrypted Data Upload.} 
In this module, a seller registers on SSMC and the service provider stores encrypted data (cf. Fig.\ref{fig:combined}.a). 

\noindent\hangindent 1em \textit{Step1}. When a seller expects to sell data for profits, he should first divide data into several pieces and encrypt them separately with different keys (denoted as $K_i$, where $i=1,2,...,n$), which is generated based on $K$. Such pieces of data should be valuable so that others can judge the quality of full data with the received segments. Here, $D_i=Enc^{AES}_{K_i}(Data_i)$ is the encrypted data.  

\noindent\hangindent 1em \textit{Step2}. The seller sends a registration demand in the form of a transaction. The registration demand includes the seller's information and data description. The seller information consists of $A_{seller}$ and $IP_{seller}$. Data description includes four main parts: \textit{content description}, \textit{data size}, \textit{the root} $r_d$ and \textit{the root} $r_{ed}$. Here, $r_d$ is the root of $M_d$ and $r_{ed}$ is the root of $M_{ed}$, where $M_d=\mathsf{Mtree}(Data_1,Data_2,...,Data_n)$ and $M_{ed}=\mathsf{Mtree}(D_1,D_2,...,D_n)$. They will be recorded in SSMC. Tokens will also be sent as deposit in this step and may be lost later if the data is found resold. SSMC will reject the request if the corresponding $r_d$ is the same as that of data recorded before. This mechanism prevents reselling on the blockchain platform. 

\noindent\hangindent 1em \textit{Step3,4}. After approving the seller's registration demand, SSMC stores useful information. Blockchain generates the hash of the next block and uses it as a public random $seed$.  

\noindent\hangindent 1em \textit{Step5,6}. The seller runs $\mathsf{Rand}(seed)$ to get a sequence of random numbers $I_{rand}$. The number of random numbers generated is the number of data units that need to be exposed. We assume that this number can support semantic comparison with the data description and data plagiarism detection without disclosing too much plaintext data.
The seller provides ($Data_{I_{rand}},P_d,P_{ed})$ to SSMC, where $P_d=\mathsf{Mproof}(M_d,I_{rand})$ and $P_{ed}=\mathsf{Mproof}(M_{ed},I_{rand})$. The contract SSMC then checks the operation $\mathsf{Mvrfy}(i,r_d,Data_i,P_{d_i})==1$ and $\mathsf{Mvrfy}(i,r_{ed},D_i,P_{ed})==1$. If not, SSMC stops execution and returns an error. Then, the exposed pieces of data will be compared to other pieces by utilizing the uniqueness index. Data plagiarism will result in deposit loss, preventing the reselling behavior. The authenticated data will be assigned an ID.

\noindent\hangindent 1em \textit{Step7,8}. The SP registration demand is parsed into $IP_{sp}$, $A_{sp}$, $ID_{data}$, $unit \, price$. 

\noindent\hangindent 1em \textit{Step9,10,11}. The seller sends encrypted data and Merkle proof to the service provider according to $IP_{sp}$ and confirms the registration demand so that the corresponding service provider can participate in the next stage.

\smallskip
\noindent\textbf{Matching and Data Downloading.}
In this module, a consumer registers on SCMC and selects the service provider to download data (see Fig.\ref{fig:combined}.b).

\noindent\hangindent 1em  \textit{Step1,2}. The consumer queries for data description and the exposed pieces of data. The consumer compares the description with exposed data content and selects data once receiving feedback from SSMC. 
    
\noindent\hangindent 1em  \textit{Step3,4,5}. The consumer stores the tuple $(IP_{sp},A_{cm},ID_{data})$ on SCMC and sends enough tokens to pay for the download service. These tokens will be sent to CPC and, if unfortunately the service provider or seller cheats on this transaction, will be returned to the consumer. When receiving the demand, SCMC queries SSMC with $ID_{data}$  to obtain $price$, $data size$ and $unit \, price$. Then, SCMC will verify $Tkn_{cm} \geq price+size*unit\, price$. Failed transactions will be discarded while the rest being broadcast. The seller can determine data and service providers by giving index $i$ and corresponding addresses.
    
\noindent\hangindent 1em  \textit{Step6,7,8}. The consumer contacts the service provider based on $IP_{sp}$, received in \textit{Step2}.  In \textit{Step7}, a service provider encrypts data $D$ with the random key $K_{sp}$. The service provider will calculate the Merkle result $M_{eed}$, where $M_{eed}=\mathsf{Mtree}(DD_1,DD_2,...,DD_n)$, with the Merkle root $r_{eed}$ and upload $P_{eed_i}$ to SCMC, where $P_{eed_i}=\mathsf{Mproof}(M_{eed},i)$ and $i$ is the index.
    
\noindent\hangindent 1em  \textit{Step9,10}. The selected service provider information is provided. It is composed of $A_{sp}$ and the index of downloading pieces from service providers. The consumer can download data from multiple providers for efficiency. The service provider sends $DD=Enc^{AES}_{K_{sp}}{D}$ to consumers.
    
\noindent\hangindent 1em  \textit{Step11,12}.The consumers need to verify whether $\mathsf{Mvrfy}(i,r_{eed},DD,P_{eed_i})==1$. If not, the (double-)encrypted data will be considered as an error if it cannot pass the verification, and the consumer, as a result, will not execute \textit{step14}. 
    
\noindent\textbf{Decryption and Appealing.}
In this module, the consumer pays both the service provider and the seller.

Payment to the service provider involves the following steps.(see Fig.\ref{fig:combined}.c)

\noindent\hangindent 1em  \textit{Step1,2,3}. SSMC transfers tokens and $(A_{cm}, A_{sp}, ID_{data})$ to CPC. The consumer generates a  key pair $(Pub_{cm}, Pri_{cm})$ and broadcasts $Pub_{cm}$ to CPC. CPC waits for the service provider to get $Enc_{Pub_{cm}}(K_{sp})$. 
    
\noindent\hangindent 1em  \textit{Step4,5,6}. The consumer obtains $Enc_{Pub_{cm}}(K_{sp})$ from CPC. Then, he decrypts data with $K_{sp}$ to get $D^{'}_i$. If {$\mathsf{Mvrfy(i,r_{ed},D^{'}_i,P_{ed})\neq 1}$, the consumer executes the appealing phase. Appeal contains $(Pri_{cm},i, DD_i)$. Here, $Pri_{cm}$ is generated in every download process. Otherwise, it indicates the decryption key and encrypted data received by the consumer are true, and CPC will send tokens to the service provider directly. 

\noindent\hangindent 1em  \textit{Step7,8,9}. CPC calculates $K_{sp}$ and $D^{'}_i$, where $D^{'}_i=Dec^{AES}_{K_{sp}}(DD)$ while the decryption key $K_{sp} = Dec_{pri_{cm}}(Enc_{pub_{cm}}(K_{sp}))$. Then, CPC verifies whether $\mathsf{Mvrfy}(i,r_{ed},D^{'}_i,P_{ed})\neq 1$. If it passes the verification, CPC withdraws the tokens to SSMC. Otherwise, CPC will pay the service providers.

\noindent Paying the seller is similar to paying the service providers, the differences between mainly concentrate on \textit{Step2}, \textit{Step3}, \textit{Step4}, \textit{Step7}, and \textit{Step8} (see Fig.\ref{fig:combined}.d). 

\noindent\hangindent 1em  \textit{Step2,3,4}. The consumer generates a new public-private key pair $(Pub_{cm},Pri_{cm})$  and broadcasts $Pub_{cm}$ to CPC. After listening to CPC to get $Pub_{cm}$, the seller calculates $Enc_{pub_{cm}}(K_{seller})$ and sends it to CPC. 
    
\noindent\hangindent 1em  \textit{Step7,8}. During the appealing phase, the consumer relies on his private key to prove his ownership. CPC verifies the encryption of the corresponding data, which is similar to the step of paying for service providers. The verification will determine the token flow. 

\begin{figure*}[!hbt]
    \vspace{-0.2in}
    \centering
    \includegraphics[width=\linewidth]{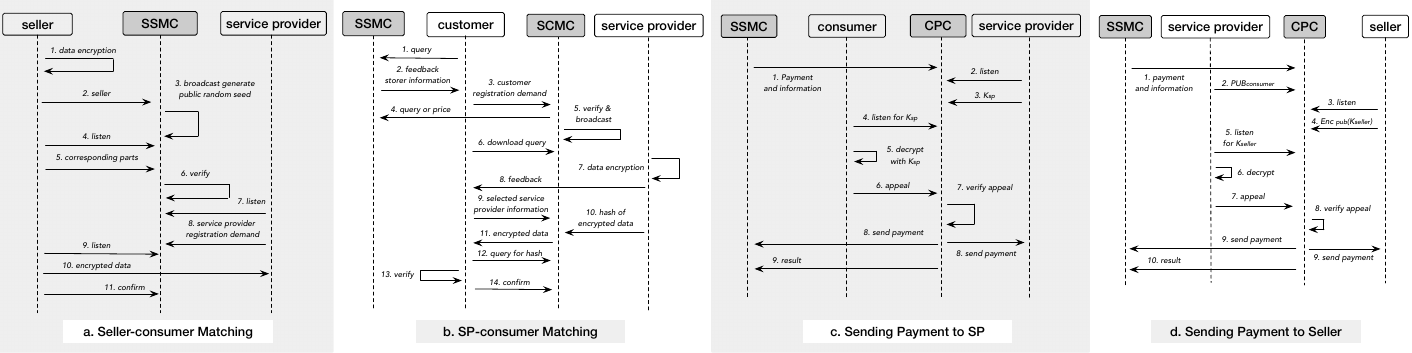}
    \setlength{\abovecaptionskip}{-0.2cm}
   \caption{Component Workflow}
    \label{fig:combined}
    \vspace{-0.3in}
\end{figure*}

\section{Security Analysis}
\label{sec-security}

In this section, we provide the analysis of BDTS based on game theory. The basic model of our solution is a \textit{dynamically sequential game} with \textit{multiple players}. The analyses are based on \textit{backward induction}. We prove that our model can achieve a subgame perfect Nash equilibrium (SPNE) if all participants honestly behave. 

Specifically, our proposed scheme consists of three types of parties, including seller (SL), service provider (SP), and consumer (CM) as shown in Fig.\ref{fig-game}. These parties will act one by one, forming a sequential game. The following party can learn the actions from the previous. Specifically, A SL will first upload the data with the corresponding encryption key to the SP (workflow in \textit{black} line). Once receiving data, the SP encrypts data by his private key and stores the raw data locally while related information is on-chain. CM searches online to find products and pay for the favorite ones both to SP and SL via smart contracts (in \textit{blue} line). Last, the SP sends the raw data and related keys to CM  (in \textit{brown} line). Based on that, we define our analysis model as follows.

\begin{figure}[!hbt]
\vspace{-0.2in}
    \centering
    \includegraphics[width=0.8\linewidth]{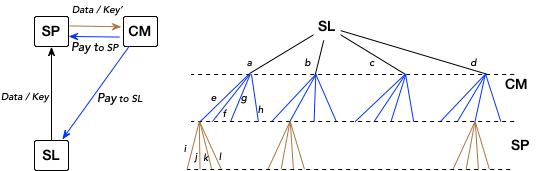}
   \caption{Game and Game Tree}
    \label{fig-game}
    \vspace{-0.3in}
\end{figure}

\begin{definition} SM-SP-CM involved system forms an extensive game denoted by  
\setlength{\abovedisplayskip}{3pt}\setlength{\belowdisplayskip}{3pt}$$\mathcal{G}=\{ \mathcal{N},\mathcal{H},\mathcal{R},P,u_i \}.$$
Here, $\mathcal{N}$ represents the participated players where $\mathcal{N}=\{SL,SP,CM\}$; $\mathcal{R}$ is the strategy set; $\mathcal{H}$ is the history, $P$ is the player function where $P: \mathcal{N}\times\mathcal{R} \xrightarrow{} \mathcal{H}$; and $u_i$ is the payoff function.
\end{definition}

Each of the participating parties, they have four strategies as defined in \textcolor{magenta}{Sec.\ref{subsec-secritymodel}} (\textit{security model}). SL has actions on both updated data and related decryption keys (AES for raw data), forming his strategies $\mathcal{R}_{SL}$, where $\mathcal{R}_{SL}=\{a,b,c,d\}$. Similarly, CM has strategies $\mathcal{R}_{CM}=\{e,f,g,h\}$ to show his actions on payments to SL or SP. SP has strategies $\mathcal{R}_{SP}=\{i,j,k,l\}$ for actions on downloading data and related keys. We list them at Table.\ref{tab:game}. However, it is not enough for quantitative analysis of the costs of these actions to be unknown. According to the market prices and operation cost, we suppose that a piece of raw data worth $10$ \textit{unit}s, while generating keys compensates $1$  \textit{unit}. The service fee during the transactions is $1$  \textit{unit}s for each party. Thus, we provide the cost of each strategy in the Table.\ref{tab:game}. The parameters of $x$ and $y$ are actual payments from CM, where $0\leq x<20,0\leq y<4, x+y<24$.
\begin{wraptable}{r}{7cm} 
\vspace{-0.15in}
\caption{Strategies and Costs: i. The cost of -11 \textit{unit}s are short for -11, applicable to all; ii. Data is sold at 20 (to SL) while the service fee is 4 (to SP).}\label{tab:game}
\vspace{-0.15in}
\begin{center}
\resizebox{1\linewidth}{!}{
\begin{tabular}{c|cc} 
 \toprule
 \multicolumn{1}{c}{\textit{\textbf{SL Strategy}} }    &  \multicolumn{1}{|c}{Matched data}          & {Non-matched data} \\ 
 \midrule
  {Matched key}        & a, -11  & b, -1 \\ 
  {Non-matched key}    & c, -10  &  d, 0\\ 
  \midrule
 \multicolumn{1}{c}{\textit{\textbf{CM Strategy}} }    &  \multicolumn{1}{|c}{Sufficiently}       & {Insufficiently} (to SL) \\ 
 \midrule
  {Sufficiently} (to SP)  & e, -24  & f, -(x+4) \\ 
  {Insufficiently Paid}   &  g, -(y+20)    &  h, -(x+y) \\ 
\midrule
 \multicolumn{1}{c}{\textit{\textbf{SP Strategy}} }    &  \multicolumn{1}{|c}{Authentic data}       &  Non-authentic data  \\ 
 \midrule
  {Matched key}       & i, -2   & j, -1 \\ 
  {Non-matched key}  &  k, -1     & l, 0   \\ 
 \bottomrule
\end{tabular}
}\end{center}
\vspace{-0.3in}
\end{wraptable}

Then, we dive into the history set $\mathcal{H}$ that reflects the conducted strategies from all parties before. For instance, the history $aei$ represents all parties performing honestly. There are 64 possible combinations (calculated by $64=4*4*4$) based on sequential steps of SL, SM, and SP. We provide their game tree in Table \ref{tab:game}. We omit their detailed representation due to their intuitive induction. Our analysis is based on these fundamental definitions and knowledge. We separately show the optimal strategy (with maximum rewards) for each party, and then show how to reach a subgame perfect Nash equilibrium, which is also the Nash equilibrium of the entire game. Before diving into the details of calculating each subgame, we first drive a series of lemmas as follows.

\begin{lemma}\label{lma-seller}
\vspace{-0.05in}
If one seller provides data not correspond to the description, the seller cannot obtain payments. 
\end{lemma}

\begin{proof}
\vspace{-0.15in}
The description and Merkle root of data are first broadcast before the generation of random indexes. Once completing the registration of the seller, the blockchain generates a random index. Exposed pieces are required to match the Merkle roots so that the seller cannot provide fake ones. Meanwhile, these pieces ensure that data can conform to the description. Otherwise, consumers will not pay for the content, and service providers will not store it, either. \qed
\end{proof}

\begin{lemma}
\label{lma-sellerkey}
\vspace{-0.05in}
If one seller provides a decryption key not conforming to the description, the seller cannot obtain payments. 
\end{lemma}

\begin{proof}
\vspace{-0.15in}
The seller encrypts data (segmented data included) with his private keys. The results of both encryption and related evidence will be recorded by the smart contract, which covers the Merkle root of encrypted data and the Merkle root of data. If a seller provides a mismatched key, the consumer cannot decrypt the data and he has to start the appeal process. As $D_i$ and receipt are owned by the consumer, if the consumer cannot obtain correct data, the consumer can appeal with evidence. The smart contract can automatically judge this appeal. If the submitted evidence is correct and decryption results cannot match the Merkle root of data, the contract will return deposited tokens to the consumer.  \qed
\end{proof}


\begin{lemma}\label{lma-cm}
\vspace{-0.05in}
A consumer without sufficient payments cannot normally use data. 
\end{lemma}

\begin{proof}
\vspace{-0.13in}
The consumer will first send enough tokens to SCMC and this code of the smart contract is safe. The smart contract will verify whether the received tokens are enough for the purchase. After the seller and consumer provide their decryption key through the smart contract, the consumer can appeal at a certain time, or it’s considered that the key is correct and payments will be distributed to the seller and service providers.  \qed
\end{proof}

\begin{lemma}
\vspace{-0.05in}
If one service provider provides data not conforming to that of the seller, he cannot obtain payments. 
\end{lemma}

\begin{proof} 
\vspace{-0.13in}
This proof is similar to Lemma \ref{lma-seller}.  \qed
\end{proof}

\begin{lemma}\label{lma-sp}
\vspace{-0.05in}
If one service provider provides a decryption key not conform to data, he cannot obtain payments. 
\vspace{-0.1in}
\end{lemma}

\noindent Here, \textbf{Lemma \ref{lma-seller}} to \textbf{Lemma \ref{lma-sp}} prove the payoff function of each behavior. Based on such analyses, we can precisely calculate the payoff function of combined strategies in our sequential game. As discussed before, a total of 64 possible combinations exist, and we accordingly calculate the corresponding profits as presented in Table \ref{tab:net}. We demonstrate that the system can reach the subgame perfect Nash equilibrium under the following theorem.

\begin{table}[!hbt]
\vspace{-0.3in}
\setlength{\abovecaptionskip}{-0.2cm}
\caption{Payoff Function and Profits (blue texts reach Nash Equilibrium)}\label{tab:net}
\begin{center}
\resizebox{1\textwidth}{!}{
\begin{tabular}{c|c|c|c|c|c|c|c} 
 \toprule
 \multicolumn{1}{c}{\textit{\textbf{$\mathcal{H}$}} }  & \multicolumn{4}{c}{\textbf{Payoff} in the form of (SL,CM,SP)} \\
 \midrule
 \textit{aei} & \textcolor{blue}{\text{(9,-4,2)}} & \textit{bei} & (19,-24,2) &  \textit{cei} & (10,-24,2) & \textit{dei} & (20,-24,2) \\
 \textit{aej} & (9,-24,3) & \textit{bej} & (19,-24,3) &  \textit{cej}& (10,-24,3) &  \textit{dej} & (20,-24,3) \\
 \textit{aek} & (9,-24,3) & \textit{bek} & (19,-24,3) &  \textit{cek}& (10,-24,3) &  \textit{dek} & (20,-24,3) \\
 \textit{ael} & (9,-24,4) & \textit{bel} & (19,-24,4) &  \textit{cel} & (10,-24,4) &  \textit{del} & (20,-24,4) \\
 \midrule
 \textit{afi} &  \textcolor{blue}{\text{(x-11,16-x,2)}} & \textit{bfi} & (x-1,-24,2) &  \textit{cfi}& (x-10,-24,2) &  \textit{dfi} & (x,-24,2) \\
 \textit{afj} & (x-11,-24,3) & \textit{bfj} & (x-1,-24,3) &  \textit{cfj}& (x-10,-24,3)  &  \textit{dfj} & (x,-24,3) \\
 \textit{afk} & (x-11,-24,3) & \textit{bfk} & (x-1,-24,3) &  \textit{cfk}& (x-10,-24,3) &  \textit{dfk} & (x,-24,3) \\
 \textit{afl} & (x-11,-24,4) & \textit{bfl} & (x-1,-24,4) &  \textit{cfl}& (x-10,-24,4) &  \textit{dfl} & (x,-24,4) \\
 \midrule
 \textit{agi} &  \textcolor{blue}{\text{(9,-y,y-2)}} & \textit{bgi} & (19,-24,y-2) &  \textit{cgi}& (10,-24,y-2) &  \textit{dgi} & (20,-24,y-2) \\
 \textit{agj} & (9,-24,y-1) & \textit{bgj} & (19,-24,y-1) &  \textit{cgj}& (10,-24,y-1) &  \textit{dgj} & (20,-24,y-1) \\
 \textit{agk} & (9,-24,y-1) & \textit{bgk} & (19,-24,y-1) &  \textit{cgk}& (10,-24,y-1) &  \textit{dgk} & (20,-24,y-1) \\
 \textit{agl} & (9,-24,y) & \textit{bgl} & (19,-24,y) &  \textit{cgl}& (10,-24,y) &  \textit{dgl} & (20,-24,y) \\
 \midrule
 \textit{ahi} &(x-11,-x-y,y-2) & \textit{bhi} & (x-1,-24,y-2) &  \textit{chi}& (x-10,,-24,y-2) &  \textit{dhi} & (x,-24,y-2) \\
 \textit{ahj} & (x-11,-24,y-1) & \textit{bhj} & (x-1,-24,y-1) &  \textit{chj}& (x-10,,-24,y-1) &  \textit{dhj} & (x,-24,y-1) \\
 \textit{ahk} & (x-11,-24,y-1) & \textit{bhk} & (x-1,-24,y-1) &  \textit{chk}& (x-10,,-24,y-1) &  \textit{dhk} & (x,-24,y-1) \\
 \textit{ahl} & (x-11,-24,y) & \textit{bhl} & (x-1,-24,y) &  \textit{chl}& (x-10,,-24,y) &  \textit{dhl} & (x,-24,y) \\
 \bottomrule
\end{tabular}
}
\end{center}
\vspace{-0.4in}
\end{table}

\begin{theorem}
\vspace{-0.1in}
The game will achieve the only subgame perfect Nash equilibrium (SPNE) if all three parties act honestly: sellers upload the matched data and matched key, service providers adopt the authentic data, and matched decryption key, and consumers purchase with sufficient payments. Meanwhile, the SPE is also the optimal strategy for the entire system as a Nash Equilibrium. \qed
\end{theorem}

\begin{proof} \vspace{-0.1in}First, we dive into the rewards of each role, investigating their payoffs under different strategies. For the seller, we observe that the system is not stable (cannot reach Nash equilibrium) under his optimal strategies. As shown in Tab.\ref{tab:net}, the optimal strategies for sellers (\textit{dei},\textit{dej},\textit{dek},\textit{del},\textit{dgi},\textit{dgj},\textit{dgk},\textit{dgl}) is to provide mismatched keys and data, while at the same time obtain payments from consumers. However, based on Lemma \ref{lma-seller} and Lemma \ref{lma-sellerkey}, the seller in such cases cannot obtain payments due to the punishment from smart contracts. These are impractical strategies when launching the backward induction for the subgame tree in Fig.\ref{fig-game}. Similarly, for both consumers and service providers, the system is not stable and cannot reach Nash equilibrium under their optimal strategies. Based on that, we find that the optimal strategy for each party is not the optimal strategy for the system. 
 
Then, we focus on strategies with the highest payoffs (equiv. utilities). As illustrated in Tab.\ref{tab:net} (red background), the strategies of \textit{aei}, \textit{afi} and \textit{agi} hold the maximal payoffs where $u_{aei}=u_{afi}=u_{agi}=7$. Their payoffs are greater than all competitive strategies in the history set $\mathcal{H}$. This means the system reaches Nash equilibrium under these three strategies. However, multiple Nash equilibriums cannot drive the most optimal strategy because some of them are impractical. 

We conduct the backward induction for each game with Nash equilibriums. We find that only one of them is the subgame perfect Nash equilibrium with feasibility in the real world. Based on Lemma \ref{lma-cm}, a consumer without sufficient payments, either to the seller or service provider, cannot successfully decrypt the raw data. He will lose all the paid money ($x+y$). This means both \textit{afi} and \textit{agi} are impractical. With the previous analyses in the arm, we finally conclude that only the strategy \textit{aei}, in which all parties act honestly, can reach the subgame perfect Nash equilibrium. This strategy is also the Nash equilibrium for the entire BDTS game.  \qed
\end{proof}

\section{Implementation and Evaluation}
\label{sec-efficiency}


\noindent\textbf{Implementation and Configurations.} We provide the detailed implementation of three major functions, including \textit{sharding encryption} that splits a full message into several pieces, \textit{product matching} to show the progress of finding a targeted product, and \textit{payment} that present the ways to pay for each participant. Our full practical implementation is based on Go language with 5 files, realizing the major functions of each contact that can be operated on Hyperledger platform\footnote{https://github.com/YXJpYQ/BDTS\_Blockchain\_based\_Data\_Trading\_System.git}. We provide implementation details in Appendix A. 

Our evaluation operates on Hyperledger Fabric blockchain \cite{androulaki2018hyperledger}, running on a desk server with Intel(R) Core(TM) i7-7500U CPU\@2.70GHz and 8.00 GB RAM. We simulate each role (\textit{consumer},  \textit{seller} and \textit{service providers}) at three virtual nodes, respectively. These nodes are enclosed inside separated dockers under the Ubuntu 18.04 TLS operating system. 

\smallskip
\noindent\textbf{Computational Complexity.} Firstly, we provide a theoretical analysis of computational complexity and make comparisons with competitive schemes. We set $\tau_E$, $\tau_{E_{A}}$, $\tau_D$, $\tau_{D_{A}}$, $\tau_M$ and $\tau_V$ to separately represent the asymmetric encryption time, symmetric encryption (AES) time, asymmetric decryption time and symmetric encryption time, the Merkle tree merging operation time and the Merkle proof verification time. We give our theoretical analysis of each step in Tab.\ref{tab-complexity}.

Firstly, at the \textit{encrypted data uploading} module, the seller will divide the entire data into several pieces of data and upload their proofs on-chain. We assume the data has been split into $n$ pieces, and every piece of data $Data_i$ needs to be encrypted into $D_i$. Then, these encrypted data have been stored at the Merkle leaves, merging both $Data_i$ and $D_i$ to obtain $M_d$ and $r_{ed}$.  Secondly, at the \textit{matching and data downloading} module, the consumer can select service providers to download different data segments from them. Before providing the service, the service provider needs to encrypt the received $D_i$ with their private keys, accompanied by corresponding Merkle proofs as in the previous step. Here, the encryption is based on a symmetric encryption algorithm. Once completed, multiple downloads occur at the same time. More service providers will improve the efficiency of downloading because the P2P connection can make full use of network speed. Last, at the \textit{decryption and appealing} module, the consumer obtains each encrypted piece of data and starts to decryption them. They need to verify whether the received data and its proof are matched. If all pass, they can use the valid keys (after payment) for the decryption. Here, the appeal time is related to the number of appeal parts instead of the appeal size.  

We further make a comparison, in terms of on-chain costs, with existing blockchain-based fair exchange protocols. Gringotts \cite{goyal2019secure} spends $O(n)$ as they store all the chunks of delivering data on-chain. CacheCas \cite{almashaqbeh2019cachecash} takes the cost at a range of  $[\mathcal{O}(1), \mathcal{O}(n)]$ due to its \textit{lottery tickets} mechanism. FairDwonload \cite{he2021fair}, as they claimed, spends  $\mathcal{O}(1)$. But they separate the functions of delivering streaming content and download chunks. Our protocol retains these functions without compromising efficiency, which only takes  $\mathcal{O}(1)$.

\begin{table}[!hbt]
  \vspace{-0.35in}
  \renewcommand\arraystretch{1.1}
\caption{Computational Complexity and Comparison ($i$ is the number of segmented data;  $n$ represents a full chunk of data)} 
 \label{tab-complexity}
 \belowrulesep=0pt
 \aboverulesep=0pt 
 \resizebox{\linewidth}{!}{
 \begin{tabular}{c|cc|c}
    \toprule
     \multicolumn{1}{c}{\textbf{Algorithm}}   & \multicolumn{1}{c}{\textbf{Complexity}}  
      & \multicolumn{1}{c}{\textbf{Schemes}}
      & \multicolumn{1}{c}{\textbf{On-chain Cost}}\\ 
     \cmidrule(r){1-2}
     \cmidrule{3-4}
   Encrypted data uploading & $i{\tau_E}+2{\tau_M}+2{\tau_V}$   &  \multicolumn{1}{c|}{Gringotts \cite{goyal2019secure} } &   $\mathcal{O}(n)$ \\
   
   Matching and Data downloading & $i{\tau_{E_{A}}}+2{\tau_M}+2{\tau_V}$ & \multicolumn{1}{c|}{ CacheCash \cite{almashaqbeh2019cachecash} } &   $[\mathcal{O}(1), \mathcal{O}(n)]$  \\
   
   Encryption and appealing &   $ i{\tau_D+\tau_{D_{A}}}+2{\tau_M}+2{\tau_V}$  &  \multicolumn{1}{c|}{ FairDwonload \cite{he2021fair} } &  $\mathcal{O}(1)$  \\
    \cmidrule(r){1-2}
 \multicolumn{2}{c}{} &  \multicolumn{1}{c|}{\textbf{\textit{BDTS(Ours)}}}  &   $\mathcal{O}(1)$   \\
   \cmidrule(l){3-4}
  \end{tabular}
  }  
  \vspace{-0.2in}
\end{table}

\smallskip
\noindent\textbf{Efficiency.} Then, we launch experimental tests to evaluate efficiency in multi-dimensions. We focus on the \textit{download} functionaries, the most essential function (due to high frequency \& large bandwidth) invoked by users. 

\textit{\textbf{Data Type.}} We evaluate three mainstream data types, covering text, image, and video. The text-based file is the most prevailing data format in personal computers. As a standard, a text file contains plain text that can be edited in any word-processing program. The image format encompasses a variety of different subtypes such as TIFF, png, jpeg, and BMP, which are used for multiple scenarios like printing or web graphics (e.g., NFT \cite{wang2021non}). We omit subtle differences between each sub-format because they perform equivalently in terms of download services. Similarly, video has a lot of sub-types including MP4, MOV, MP4, WMV, AVI, FLV, etc. We only focus on its general type. From the results in Fig.\ref{fig-tests}, we can observe that all three types of data have approximately the same performance, under different configurations of data size and storage capacity. The results indicate that \textit{the performance of the download service has no significant relationship with the data type.} This is an intuitive outcome that can be proved by our common sense. The upload/download service merely opens a channel for inside data, regardless of its content and types. This also shows that BDTS can support multiple types of data without compromising efficiency.

\textit{\textbf{Data Size.}} We adjust data sizes at three levels, including 10M, 100M, and 1G, to represent a wide range of applications at each level. As shown in Fig.\ref{fig-tests}, 10M data (Text, 1 storage) costs at most no more than $2$ seconds, 100M data in the same format spends around 18s, and 1G data costs 170s. The results indicate that \textit{the download time is positively proportional to its data size.} The larger the data, the slower it downloads. This can also apply to different types of data and different storage capacities. A useful inspiration from evaluations of data size is to ensure a small size. This is also a major consideration to explain the reasons for splitting data into pieces in our BDTS. The splitting procedure can significantly improve service quality either for uploading or downloading. Sharded data can be reassembled into its full version once all pieces of segments.

\begin{figure*}[!hbt]
    \vspace{-0.15in}
    \centering
    \includegraphics[width=\linewidth]{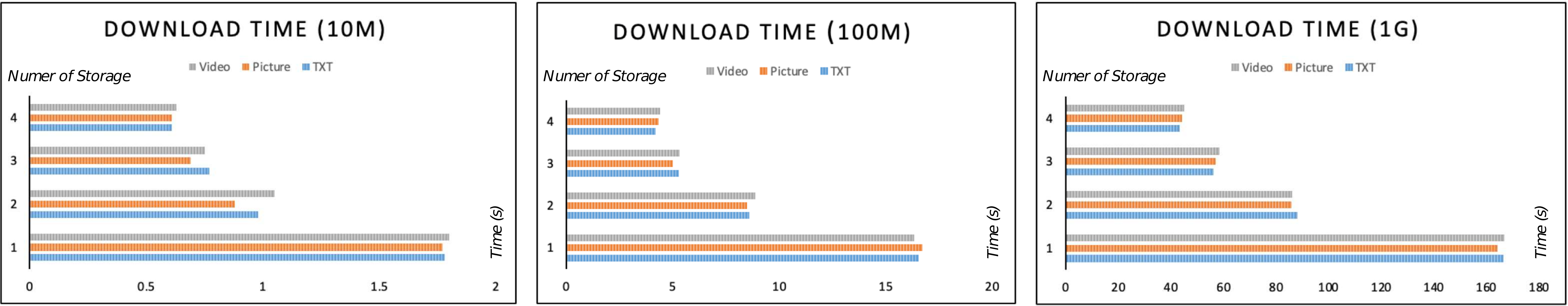}
    \setlength{\abovecaptionskip}{-0.3cm}
   \caption{Download Times of Different Data Type, Data Size and Storage Capacity: We evaluate three types of data formats including video \textit{(grey)}, image \textit{(orange)}, and text \textit{(blue)}. For each type, we test download times in distinguished data size with 10M \textit{(left)}, 100M \textit{(middle)}, and 1G \textit{(right)}. Meanwhile, we also investigate the performance along with an increased number of storage devices \textit{(from 1 to 4)}, or equiv. the number of service providers.}
    \label{fig-tests}
 \vspace{-0.25in}
\end{figure*}

\smallskip
\textit{\textbf{Storage Capacity.}} The storage capacity refers to the number of storage devices that can provide download services. The device is a general term that can be a single laptop or a cluster of cloud servers. If each service provider maintains one device, the number of devices is equal to the number of participating service providers. We adjust the storage capacity from 1 device to 4 devices in each data type and data size. All the subfigures (the columns in \textit{left}, \textit{middle} and \textit{right}) in Fig.\ref{fig-tests} show the same trend: \textit{increasing the storage capacity over the distributed network will shorten the download time.} The result can apply to all the data types and data sizes. The most obvious change in this series of experiments is adding devices from 1 to 2, which is almost short half of the download time. A reasonable explanation might be that a single-point service is easily affected by other factors such as network connection, bandwidth usage, or propagating latency. Any changes in these factors may greatly influence the download service from users. But once adding another device, the risk of single-point diminishes as the download service becomes decentralized and robust. More connections can drive better availability, as also proved by setting devices to 2, 3 and 4. This is why BDTS allows consumers to download data from multiple providers.

\textit{\textbf{Average Time.}} We dive into one of the data types to evaluate its i) average download times that are measured in MB/sec by repeating multiple times of experiments under different data sizes; and ii) the trend along with the increased number of storage devices. Compared to previous evaluations, this series of experiments scrutinize the subtle variations under different configurations, figuring a suite of curves. As stated in Tab.\ref{tab-avgtime}, the average downloading times under the storage capacity (from 1 to 6) are respectively $0.167$s, $0.102$s, $0.068$s, $0.051$s, $0.039$s, and $0.031$s. Their changes start to deteriorate, approaching a convex (downward) function as illustrated in 
Fig.\ref{fig-performance}. This indicates that the trend of download time is not strictly proportional to the changes in storage capacity. They merely have a positive relation, following a diminishing marginal effect.

\begin{figure*}
\vspace{-0.3in}
    \begin{minipage}[!h]{0.62\linewidth}
    \setlength{\abovecaptionskip}{0.cm}
    \captionof{table}{Average Download Time} 
 \label{tab-avgtime}
 \resizebox{0.93\linewidth}{!}{
 \begin{tabular}{c|cccccc|cc}
    \toprule
     \multicolumn{1}{c}{\quad}   & \multicolumn{6}{c}{\quad\textbf{Data Size (Text)}\quad}  
     & \multicolumn{1}{c}{\quad\textbf{Average Time }\quad}   \\ 
     \midrule
    \multicolumn{1}{c|}{\textbf{Storage}}   &  {\textbf{1M}}  &  {\textbf{10M}}  &  {\textbf{50M}} &  {\textbf{100M}}  &    {\textbf{500M}}    & \textbf{1G}  & 
    \quad\textbf{(s)}\quad     \\
    \midrule
   1 & 0.16 & 1.78 & 7.96 & 16.55 & 80.52 & 166.45 & 0.167   \\ 
   2 & 0.10 & 0.98 & 4.89 & 8.60 & 43.48 & 88.04  & 0.102 \\ 
   3 & 0.07 & 0.77 & 2.54 & 5.29 &  27.44   & 56.15  & 0.068  \\ 
   4 & 0.05 & 0.61 & 2.03 &  4.21 & 22.22 & 43.51  &  0.051 \\
   5 & 0.04 & 0.38 & 1.79 & 3.33  & 18.88  & 34.52  & 0.039  \\
   6 & 0.03 & 0.32 & 1.56 & 2.88 & 14.69 &   29.48  & 0.031  \\
   \bottomrule
  \end{tabular}
 }  
 \end{minipage}
 \begin{minipage}[!h]{0.33\linewidth}
 \vspace{0.2in}
    \includegraphics[height = 2.8cm]{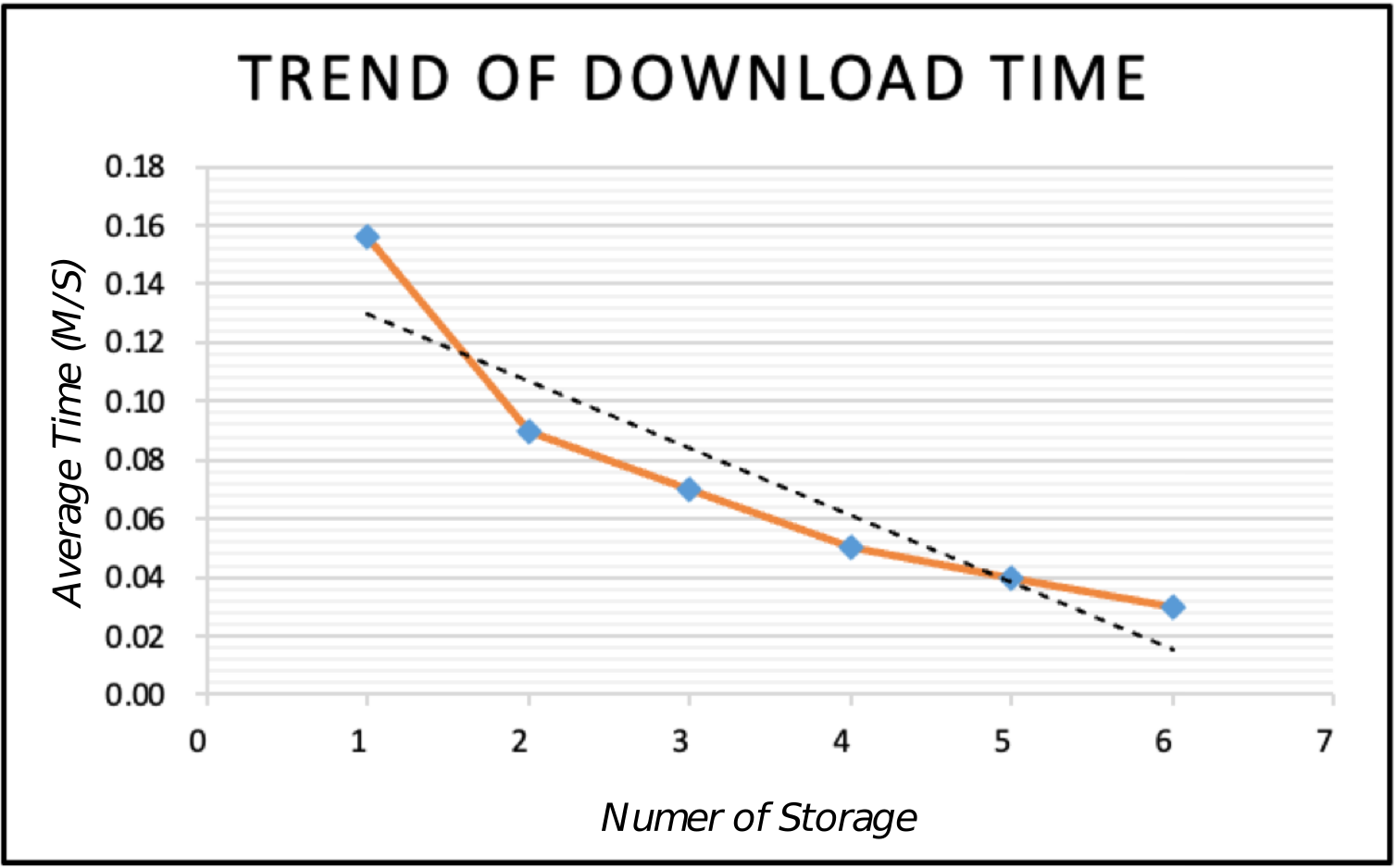}
    \setlength{\abovecaptionskip}{-0.35cm}
    \caption{Download Time}
    \label{fig-performance}
    \end{minipage}
\vspace{-0.3in}
\end{figure*}



\smallskip
\noindent\textbf{Practicability.}  We further discuss the practicality. We highlight several major features of BDTS by digging into its \textit{usability}, \textit{compatibility}, and \textit{extensibility}.

\textit{\textbf{Usability.}} Our proposed scheme improves usability in two folds. Firstly, we separately store the raw data and abstract data. The raw data provided by the sellers are stored at the local servers of service providers, while the corresponding abstract data (in the context of this paper, covering \textit{data}, \textit{description} and \textit{proof}) is recorded on-chain. A successful download requires matching both decryption keys and data proofs under the supervision of smart contracts. Secondly, the data trade in our system includes all types of streaming data such as video, audio, and text. These types can cover the largest range of existing online resources.

\textit{\textbf{Compatibility.}} Our solution can be integrated with existing crypto schemes. To avoid repeated payment, simply relying on the index technique is insufficient. The watermarking \cite{yang2020collusion} technique is a practical way to embed a specific piece of mark into data without significantly changing its functionality. It can also incorporate bio-information from users, greatly enhancing security. Beyond that, the storage (encrypted) data can leverage the hierarchical scheme \cite{gentry2002hierarchical} to manage its complicated data, as well as maintain the efficiency of the fast query.

\textit{\textbf{Extensibility.}} BDTS can extend functionalities by incorporating off-chain payment techniques (also known as layer-two solutions \cite{gudgeon2020sok}. Off-chain payment has the advantage of low transaction fees in multiple trades with the same person. Besides, existing off-chain payment solutions have many advanced properties such as privacy-preserving and concurrency \cite{malavolta2017concurrency}\cite{green2017bolt}. Our implementation only set the backbone protocol for fair exchange, leaving many flexible slots to extend functionalities by equipping matured techniques.  
\vspace{-0.1in}

\section{Conclusion}
\label{sec-conclu}

\vspace{-0.1in}
This paper explores the fairness issue in current data transaction solutions where traditional centralized authorities are not subject to any oversight due to their superpowers. Our proposed scheme, BDTS, addresses such issues by leveraging blockchain technology with well-designed smart contracts. The scheme utilizes automatically operating smart contracts to act in the role of a data executor with transparency and accountability. Our analyses, based on strict game theory induction, prove that the game can achieve a subgame perfect Nash equilibrium with optimal payoffs under the benign actions of all players. Furthermore, we implemented the scheme on the Hyperleder Fabric platform and evaluated that the system can provide users with fast and reliable service.


\vspace{-0.1in}
\bibliographystyle{unsrt}
\bibliography{bibb}
\vspace{-0.1in}

\vspace{-0.2in}
\section*{Appendix A. Implementation Details}
We give more implementation details by focusing on three major components.

\smallskip

\textit{\textbf{Sharding Encryption.}} Based on the real scenario, data transmitted in our system is large in scale.
A promising way for transferring the data is delivering them in segments (also known as \textit{data sharding}). 
Data sharding in BDTS does not affect the system consensus or consistency. Instead, data sharding is an off-chain operation conducted by sellers that will be processed before uploading. A full piece of data is split into several shards (or pieces, segments), being encrypted and stored in different memories. The blockchain only reserves its sequential orders and related evidence such as descriptions, addresses, and proofs. When a consumer confirms the purchase, he needs to download the data on service providers according to the storage list and obtain the decryption key after successful payment. Then, he can decrypt the data in pieces and finally resemble them according to the sequences for the entire piece of data. We implement the data sharding with the logic in Algm.1. Given the size of a slot (indicating the expected size of a shard), we first calculate the number of data segments (\textit{line 4}). Then, a full piece of data is split into $n$ segments (\textit{line 5}). The seller then create its encryption keys $(K_1, K_2, ..., K_n)$ based on his master private key $K_{seller}$ (\textit{line 6}). Once data splitting and key generation, the algorithm starts to encrypt each data segment under the seller's private keys (\textit{line 8-11}). Encrypted data also generates its proofs for further verifications (\textit{line 10}). Last, both raw data and encrypted data are stored on the leaves of the Merkle tree to create on-chain roots $MT1$ and $MT2$ (\textit{line 12-14}).

\textit{\textbf{Product Matching.}} It describes the process of searching for a targeted source from service providers. In the context of Algm.2, the terms \textit{keyword},  \textit{choice},  \textit{ProductList},  \textit{SPList},  \textit{MD}, \textit{Data} and \textit{Desc} represent the searchable keywords, user's preferences of products, product list, service provider list, data, and product description. When a consumer inputs a keyword, the algorithm starts to search for matched ones (\textit{line 3}) by ranging all descriptions in the product list (\textit{line 2-5}). Matched products will be recommended to a channel called \textit{showlist} for consumers. The algorithm then inputs \textit{choice} requested from the consumer, and searches related sources (encrypted data, data, description) from service providers (\textit{line 8-13}). The returned information is sent to the consumer. 

\textit{\textbf{Payment.}} This function mainly describes the method of making payments. The terms in Algm.3 \textit{cmAddr}, \textit{slrAddr}, \textit{spAddr} and \textit{Price} stand for the consumer's address, the seller's address, the service provider's address, and the price of commodities. The result (either True or False) represents the final result on whether the payment has been successfully executed. The algorithm inputs the addresses of all three entities and the commodity price (\textit{line 1}). If the token amount of the consumer is less than the selling prices, the algorithm returns false and the transaction fails (\textit{line 1-3}). Otherwise, the transaction proceeds. A major difference compared to traditional exchange protocols is that the consumer needs to pay both service providers for their on-chain services and the seller for his resources (\textit{line 4-9}).

\begin{figure*}
\vspace{-0.15in}
\subfigure{
\begin{minipage}[t]{0.32\linewidth}
\centering
\includegraphics[height=3cm]{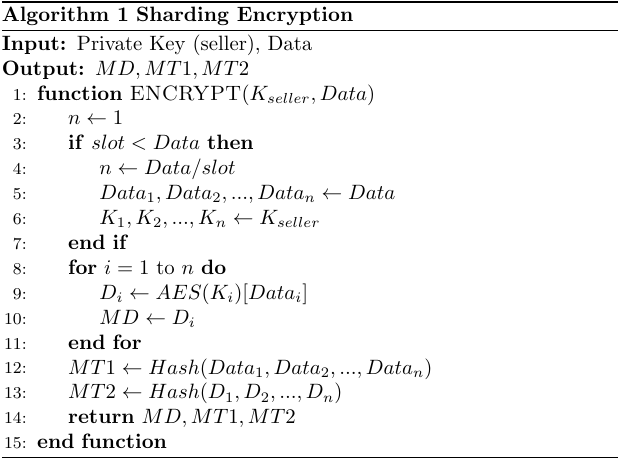}
\end{minipage}}
\subfigure{
\begin{minipage}[t]{0.31\linewidth}
\centering
\includegraphics[height=3cm]{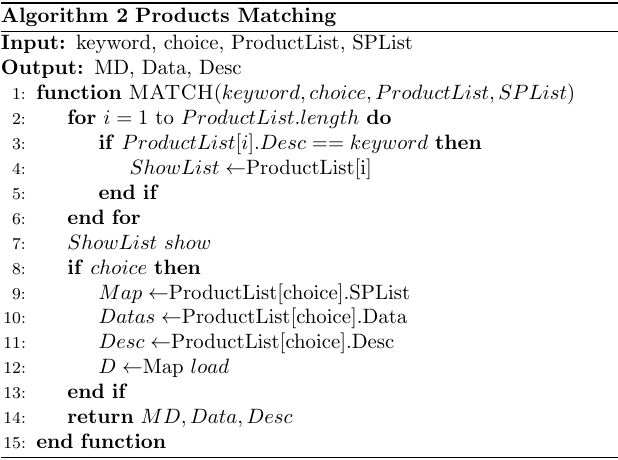}
\end{minipage}}
\subfigure{
\begin{minipage}[t]{0.3\linewidth}
\centering
\includegraphics[height=3cm]{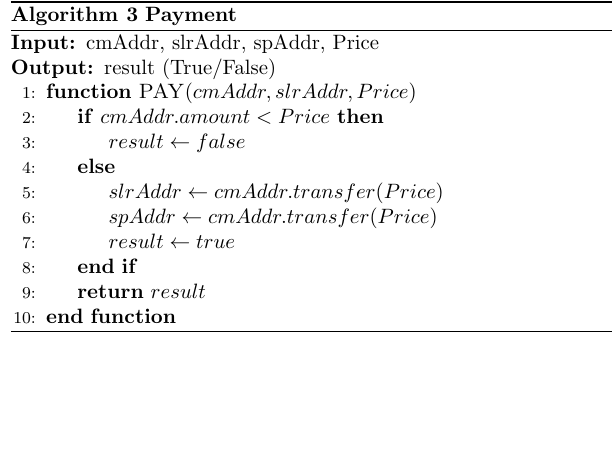}
\end{minipage}}
\caption{Major Functions}
    \label{fig:function}
\vspace{-0.2in}
\end{figure*}

\end{document}